\documentclass[final,1p,times]{elsarticle}
\usepackage{amssymb}
\usepackage{amsthm}
\usepackage{amsmath}
\usepackage{fontenc}
\usepackage[english]{babel}
\usepackage{hyperref}
\usepackage{url}
\usepackage{xurl}
\biboptions{sort&compress}

\makeatletter
\def\ps@pprintTitle{%
 \let\@oddhead\@empty
 \let\@evenhead\@empty
 \let\@oddfoot\@empty
 \let\@evenfoot\@empty
}
\makeatother

\journal{Physica A}

\begin{document}

\begin{frontmatter}

\title{Similarity networks of ordinal-pattern transitions classify falling paper trajectories}

\affiliation[inst1]{organization={Departamento de Física},
            addressline={Universidade Estadual de Maringá}, 
            city={Maringá},
            postcode={87020-900}, 
            state={PR},
            country={Brazil}}

\author[inst1]{Angelo A. Flores} 
\author[inst1]{Leonardo G. J. M. Voltarelli} 
\author[inst1]{Andre S. Sunahara} 
\author[inst1]{Haroldo~V.~Ribeiro} 
\ead{hvribeiro@uem.br}
\author[inst1]{Arthur A. B. Pessa} 
\ead{aabpessa@uem.br}

\begin{abstract}
Paper fragments in free fall constitute a simple yet paradigmatic mechanical system exhibiting remarkably complex motions. Despite a long history of investigation, this system has defied comprehensive first-principles modeling, motivating the development of phenomenological and experimental approaches to classify the free-fall dynamics of small paper fragments. Here we apply the Bandt-Pompe symbolization method to extract high-dimensional features corresponding to ordinal-pattern transitions (so-called ordinal networks) from observed area time series of video-recorded falling papers shaped as circles, squares, hexagons, and crosses. We then represent each trajectory as a node in a weighted similarity network, with edges encoding pairwise dynamical similarity, and identify motion classes via community detection. Our method automatically clusters trajectories into tumbling and chaotic falls in excellent agreement with expert visual classification. Notably, it outperforms previous approaches based on classical physical features derived from complete three-dimensional trajectories -- especially for cross-shaped papers -- and requires no prior specification of the number of motion classes. We further find that trajectories diverging from expert classifications occupy more central positions in the similarity network, suggesting more complex and ambiguous dynamic behavior.
\end{abstract}



\end{frontmatter}
\section{Introduction}\label{sec:intro}

The evolution of seemingly deterministic motions -- such as the toss of a coin or the roll of a die -- has long served as a quintessential model for randomness~\cite{blitzstein2019introduction, bartos2024fair}. These examples reveal both the challenges of mathematically modeling outwardly simple systems via equations of motion and the fundamental limits of long-term predictions under nonlinear dynamics~\cite{zeng-yuan1985onthesensitive, diaconis2007dynamical}. Similarly, the deceptively simple falling paper system has also resisted a complete first-principles description despite a long history of investigation~\cite{maxwell1853particular, tanabe1994behavior, mahadevan1995comment, field1997chaotic, belmonte1998fromflutter, mahadevan1999tumbling, pesavento2004falling, andersen2005analysis, howison2019physics, howison2020large-scale, pessa2022clustering}. Beginning with Maxwell's early qualitative characterization of tumbling paper slips~\cite {maxwell1853particular}, this system has since been studied using a wide array of theoretical and empirical methods. Classical fluid-dynamical models, for instance, typically describe this type of motion by incorporating gravitational and flow-induced torques~\cite{pesavento2004falling, andersen2005analysis}, whereas phenomenological and data-driven approaches usually focus on extracting key features of the observed dynamics~\cite{tanabe1994behavior, mahadevan1995comment, belmonte1998fromflutter, mahadevan1999tumbling, andersen2005analysis}. Indeed, in the absence of a complete analytical framework, simplified representations based on physical quantities, such as the dimensionless moment of inertia and the Reynolds number~\cite{willmarth1964steady}, have proven effective in distinguishing chaotic from regular tumbling motions~\cite{field1997chaotic}.

More recently, statistical learning has also been integrated into the methodological toolkit used to investigate the dynamics of falling papers, yielding promising results~\cite{howison2019physics, howison2020large-scale, pessa2022clustering}. For instance, Howison \textit{et al.}~\cite{howison2019physics} extensively explored functional forms capable of describing and discriminating between distinct motion patterns observed in free-falling v-shaped paper slips. In a subsequent study, Howison, Hughes, and Iida~\cite{howison2020large-scale} measured angular and vertical velocities of falling papers with various shapes to cluster steady, chaotic, and tumbling motions, as well as to study their positioning within a previously unexplored region of the parameter space defined by the moment of inertia and Reynolds number. Furthermore, Pessa, Perc, and Ribeiro~\cite{pessa2022clustering} clustered chaotic and tumbling trajectories by analyzing time series of the observed paper area via the complexity-entropy plane~\cite{rosso2007distinguishing}, constructed using two quantities derived from the Bandt and Pompe symbolization method~\cite{bandt2002permutation}. 

Considering the long history of research into falling paper dynamics and the recent methodological advancements, we introduce here the use of ordinal networks~\cite{small2013complex, mccullough2015time, pessa2019characterizing, olivares2020contrasting, pessa2020mapping} to extract vector representations of paper falling trajectories, which are subsequently combined with a network-based clustering method~\cite{lee2021non, sunahara2023universal} to systematically identify motion patterns. Specifically, our approach evaluates ordinal-pattern transitions from the observed area time series of video-recorded falling paper fragments of four distinct shapes (circles, squares, hexagons, and crosses), representing each trajectory as a node in a weighted similarity network, where edges encode pairwise dynamical similarity quantified from these ordinal-pattern transitions. We then apply the infomap algorithm~\cite{rosvall2008maps,rosvall2009map} to this network to detect communities, thereby identifying distinct classes of motion. This procedure automatically clusters free-falling paper trajectories into two groups corresponding to tumbling and chaotic motions, achieving excellent agreement with expert visual classification and outperforming previous automated approaches~\cite{howison2020large-scale, pessa2022clustering}, with the additional advantage of not requiring prior specification of the number of motion classes. Our approach particularly excels for cross-shaped fragments, further indicating that trajectories diverging from expert classifications typically occupy more central positions within the similarity network, suggesting more complex and potentially ambiguous dynamical behaviors. 

In the remainder of this paper, we first describe the dataset underlying our investigations, comprising hundreds of video-recorded free-fall trajectories of small slips of paper. We then introduce the concept of ordinal network, outline its application to the observed area time series, and detail how the information encoded in ordinal-pattern transitions is used to create a similarity network for identifying distinct classes of motion via a network community algorithm. We further compare our automatic classification with expert visual assessments and previous approaches reported in the literature. We conclude by summarizing and discussing our main findings.

\section{Data and methods}\label{sec:data}

We analyze the free-fall dataset collected by Howison, Hughes, and Iida~\cite{howison2020large-scale}, which is publicly available~\cite{howison2019github}. This database comprises video recordings of small paper slips (shaped as circles, hexagons, squares, or crosses) released from a height of $1.1$~m using a robotic arm in a controlled laboratory environment. From these recordings, Howison and colleagues~\cite{howison2020large-scale} extracted the trajectories of the centers of mass and the time series of the observed surface areas relative to the cameras capturing the motion. The trajectories were further manually classified by experts into three categories based on falling behavior: steady and periodic, tumbling, and chaotic. Steady and periodic falls are characterized by a relatively straight vertical descent of the center of mass, accompanied by periodic fluttering oscillations. Tumbling falls typically involve continuous end-over-end rotations combined with lateral drift, whereas chaotic behavior is marked by irregular alternations between tumbling and large, swooping motions without apparent regularity. 

\begin{figure*}[!ht]
    \centering
    \includegraphics[width=1\textwidth, keepaspectratio]{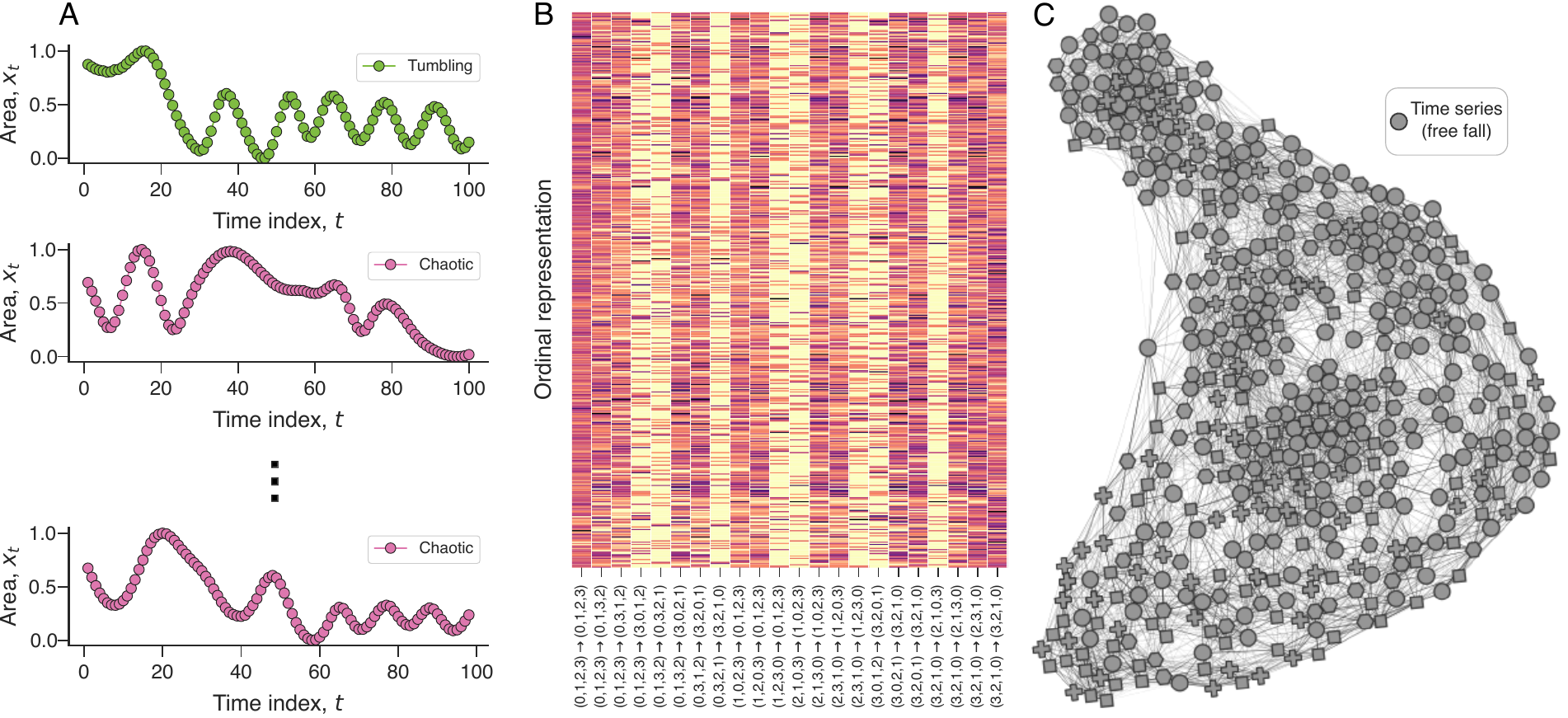}
    \caption{Clustering the dynamics of free-falling papers using ordinal-pattern transitions. (A) Examples of observed area time series $\{x_t\}_{t=1,\dots,N}$ extracted from video recordings of small paper slips in free fall, manually labeled by experts as exhibiting tumbling (green) or chaotic (pink) motion. (B) Ordinal-pattern representations of all 441 time series in the dataset. Rows in the matrix plot correspond to trajectories, whereas columns indicate the transition probability between ordinal patterns (encoded by the colors). For clarity, only a subset of the 96 possible transitions for embedding dimension $d = 4$ (those consistently exhibiting high frequencies across time series) is shown. Transition probabilities are column-wise normalized and encoded by color intensity, with darker shades indicating higher values. (C) Similarity network of trajectories constructed using a fuzzy simplicial complex derived from the uniform manifold approximation and projection (UMAP) algorithm. Nodes represent individual trajectories, with marker shapes denoting paper geometry, while weighted links (indicated by edge thickness) encode pairwise similarity.}
\label{fig:1}
\end{figure*}

Since steady motion can be readily distinguished from the other two classes of motion, and in line with previous studies~\cite{howison2020large-scale, pessa2022clustering}, we restrict our investigations to chaotic and tumbling falls, yielding 441 trajectories of papers shaped as circles (170), hexagons (120), squares (92) and crosses (59). We further focus exclusively on the observed area time series $\{x_t\}_{t=1,\dots,N}$, which are directly extracted from the video recordings and encode less information than the complete three-dimensional trajectories, thereby making the classification task, in principle, more challenging. In these time series, $x_t \approx 1$ indicates that the paper's area is fully captured by the camera, while $x_t \approx 0$ corresponds to the paper being perpendicular to the image plane captured by the camera. These series were sampled at time intervals of approximately $0.01$~s (98~Hz sampling frequency), and examples of changes in the observed area during tumbling and chaotic free-falls are shown in Figure~\ref{fig:1}A. 

The first step of our approach involves mapping the observed area time series $\{x_t\}_{t = 1,\dots,N}$ into ordinal networks~\cite{small2013complex, mccullough2015time, pessa2019characterizing, olivares2020contrasting, pessa2020mapping}. To describe this technique, we first introduce the Bandt and Pompe symbolization method~\cite{bandt2002permutation}. We begin by setting the length $d$ of a sliding window (also referred to as the embedding dimension) that traverses the series and partitions it into segments $w_{t'} = (x_{t'}, x_{t'+1},\dots,x_{t'+d-1})$ for $t' = 1,\dots,N-d+1$. For each segment $w_{t'}$, we evaluate the permutation $\pi_{t'}$ of the partition indices that arranges the corresponding values in ascending order. For example, if $\{x_t\} = \{0.5,0.7,0.6,0.4,1.0,0.4\}$ and $d = 4$, the first partition is $w_{1}= (0.5,0.7,0.6,0.4)$ with associated indices $0, 1, 2, {\rm and \ } 3$. Since $0.4 < 0.5 < 0.6 < 0.7$, the permutation that sorts the elements of $w_1$ in ascending order is $\pi_{1} = (3,0,2,1)$. In cases where equal values (draws) occur, we follow the literature and effectively treat earlier observations as smaller than later ones~\cite{cao2004detecting}. For instance, for the partition $w_{3} = (0.6, 0.4, 1.0, 0.4)$, the permutation $\pi_{3} = (1,3,0,2)$ sorts the partition in ascending order while preserving the occurrence order of the repeated elements valued at $0.4$ (indices $1$ and $3$, respectively). After evaluating the permutations of all partitions, we obtain a symbolic sequence $\{\pi_{t'}\}_{t' = 1, \dots, N-d+1}$, where each $\pi_{t'}$ belongs to the set of all $d!$ possible permutations. For $d = 4$, each $\pi_{t'}$ corresponds to one of the $d!=24$ possible permutation types $\{\Pi_i\}_{i = 1, \dots, 24}$, where $\Pi_1 = (0,1,2,3)$, $\Pi_2 = (0,1,3,2)$, and so on, up to $\Pi_{24} = (3,2,1,0)$. Using this symbolic sequence, we then compute the probability $\tilde{\rho}_{i\to j}$ of transitions between pairs of permutations $\Pi_i$ and $\Pi_j$ via
\begin{equation}\label{eq:edge_weights}
    \tilde{\rho}_{i \to j} = \frac{\text{number of transitions $\Pi_i \to \Pi_j$ in $\{\pi_{t'}\}$}}{N-d}\,.
\end{equation}
These probabilities define a graph where nodes represent the possible permutation types, and weighted directed edges among them indicate the transition frequencies -- a structure known as ordinal networks~\cite{small2013complex, mccullough2015time, pessa2019characterizing, olivares2020contrasting, pessa2020mapping}. Interestingly, because elements of adjacent partitions overlap, certain transitions are forbidden [for instance, $(0,1,2,3)$ cannot be preceded or succeeded by $(3,2,1,0)$], restricting the number of possible transitions to $d\times d!$~\cite{pessa2019characterizing}.

The embedding dimension $d$ is the only parameter that must be specified before applying this approach to our time series. Its selection, however, is not entirely arbitrary, as the number of possible permutations and transitions increases rapidly with $d$, requiring sufficiently long time series to reliably estimate each transition probability. Typically, researchers select the embedding dimension based on heuristics such as $d!\ll N$ or $5 d! \leq N$~\cite{amigo2008combinatorial}. In our case, the fall of paper slips lasts only a few seconds, resulting in relatively short time series comprising a few hundred points, which leads us to fix $d=4$ in all analyses. Consequently, each of the 441 time series of the observed area of falling papers is mapped into a set of $d\times d! = 96$ ordinal-pattern transitions, computed using the \texttt{ordpy} Python module~\cite{pessa2021ordpy}. For simplicity, we represent these probabilities as an array $(\rho_1,\rho_2,\dots,\rho_{96})$, where $\rho_1$ refers to $\tilde{\rho}_{1\to 1}$ (frequency of transitions $\Pi_{1} \to \Pi_1$), $\rho_2$ to $\tilde{\rho}_{1 \to 2}$ (frequency of transitions $\Pi_{1} \to \Pi_2$), and so on, up to $\rho_{96}$ which corresponds to $\tilde{\rho}_{24 \to 24}$ (frequency of transitions $\Pi_{24} \to \Pi_{24}$), as illustrated in Figure~\ref{fig:1}B. 

These ordinal representations capture the dynamical behavior of paper fragments in free fall by encoding the relative frequencies of local ordinal transitions along their trajectories, as inferred from the time series of observed area. These transition probabilities highlight the underlying ordering structure of the time series and remain invariant under monotonic transformations -- such as those caused by the increasing distance of a paper fragment from the camera, which typically leads to a progressive decrease in the observed area. By emphasizing temporal order rather than amplitude, this method also exhibits enhanced robustness to noise~\cite{pessa2019characterizing}, reinforcing the utility of ordinal techniques as a systematic framework for characterizing the dynamics of falling papers. Indeed, prior work~\cite{pessa2022clustering} has used an ordinal representation -- the complexity-entropy plane~\cite{rosso2007distinguishing} -- to distinguish tumbling from chaotic motions with notable success. However, that approach considers only the frequency of individual ordinal patterns, disregarding the transitions between them, and relies on the $k$-means clustering algorithm, which requires a predefined number of clusters. In contrast, our method incorporates ordinal-pattern transitions and allows the number of motion classes to emerge from the data itself.

Having obtained vector representations for each falling paper trajectory, we construct a similarity network in which nodes correspond to individual trajectories and weighted edges reflect the pairwise similarities among them. To do so, we use a recent network-based cluster approach~\cite{lee2021non, sunahara2023universal} built upon the uniform manifold approximation and projection (UMAP) algorithm -- a state-of-the-art technique for dimensionality reduction~\cite{mcinnes2018umap} -- which has produced notable results across diverse contexts, including clustering extracellular spike waveforms~\cite{lee2021non}, the analysis of scholars' productivity trajectories~\cite{sunahara2023universal}, and the characterization of electronic-game popularity series~\cite{cunha2024shape}. In this approach, the similarity network is obtained from the first step of the UMAP algorithm, which creates a fuzzy simplicial complex -- a weighted graph that captures the local topological structure of the data manifold by encoding the likelihood of connections between data points. This graph is built by identifying the $k$-nearest neighbors for each trajectory vector based on the Euclidean distance, with weights assigned via a smooth exponential kernel that accounts for local density variations. The resulting similarity network provides a high-dimensional representation that strives to preserve both the local and global structure of the data's intrinsic geometry~\cite{lause2024theart}, enabling the identification of meaningful clusters of patterns. Notably, in contrast to the second step of the UMAP algorithm, which optimizes a low-dimensional projection of the original graph by minimizing the cross-entropy between them through stochastic gradient descent, the first step is fully deterministic. In our analysis, we fix the number of nearest neighbors in the UMAP algorithm to $25$ and retain default values for all other parameters as implemented in the Python package \texttt{umap}~\cite{mcinnes2018umap-software, edler2022themap}. This setting corresponds to a moderate neighborhood size relative to our dataset of 441 trajectories and reflects a commonly adopted trade-off between local fidelity and global structure preservation; nevertheless, this choice is not critical as similar patterns emerge for alternative neighborhood sizes. Figure~\ref{fig:1}C depicts a visualization of the resulting similarity network, where nodes represent individual trajectories and edges, whose thickness encodes weight, indicate pairs of trajectories with similar vector representations derived from ordinal-pattern transitions.

The final step of our approach involves clustering the similarity network using a network community detection algorithm. Following Ref.~\cite{sunahara2023universal}, we use the Infomap algorithm~\cite{rosvall2008maps, rosvall2009map} to identify the underlying community structure, as it explicitly accounts for edge weights and consistently ranks among the best-performing methods for detecting planted partitions in benchmark networks~\cite{lancichinetti2009community, fortunato2010community, fortunato2016community}. Infomap is an information-theoretic algorithm that identifies modules in networks by modeling the flow of information as a random walk. It operates by minimizing the map equation~\cite{rosvall2008maps, rosvall2009map}, which quantifies the theoretical limit for compressing a description of such flows. The core principle is that if a random walker tends to remain within certain groups of nodes -- indicative of dense internal connections -- it assigns shorter codewords to intra-community movements and longer ones to inter-community transitions leads to a more efficient encoding of the walker's path. Infomap achieves this by constructing a two-level codebook: one for movements within communities and another for transitions between them. The algorithm iteratively refines the network partition to minimize the description length of the random walks, thereby revealing the community structure that best captures the flow dynamics. This method is particularly well-suited to our similarity network, as it naturally incorporates the edge weights reflecting the pairwise similarities among paper trajectories and models the data's intrinsic geometry through the dynamics of information flow. 

\section{Results}\label{sec:results}

The best partitioning of the network emerges from Infomap's iterative process without the need to specify the number of modules. Due to the algorithm's stochastic nature, however, repeated runs produce similar yet non-identical partitions. To account for this variability, we analyze the results of 1,000 independent realizations of the algorithm, each using a different random seed. These partitions range from two to five modules, with two-module partitions being the most frequent, occurring in nearly half ($460$) of the Infomap realizations. Notably, partitions containing two modules consistently exhibit smaller minimal description lengths than those with more modules. We therefore select this configuration as the best partitioning for our similarity network. These two-module partitions differ slightly from each other, and we select the one with the smallest minimal description length as the most consistent and parsimonious community structure of our similarity network.

\begin{figure*}[ht!]
    \centering
    \includegraphics[width=1\textwidth, keepaspectratio]{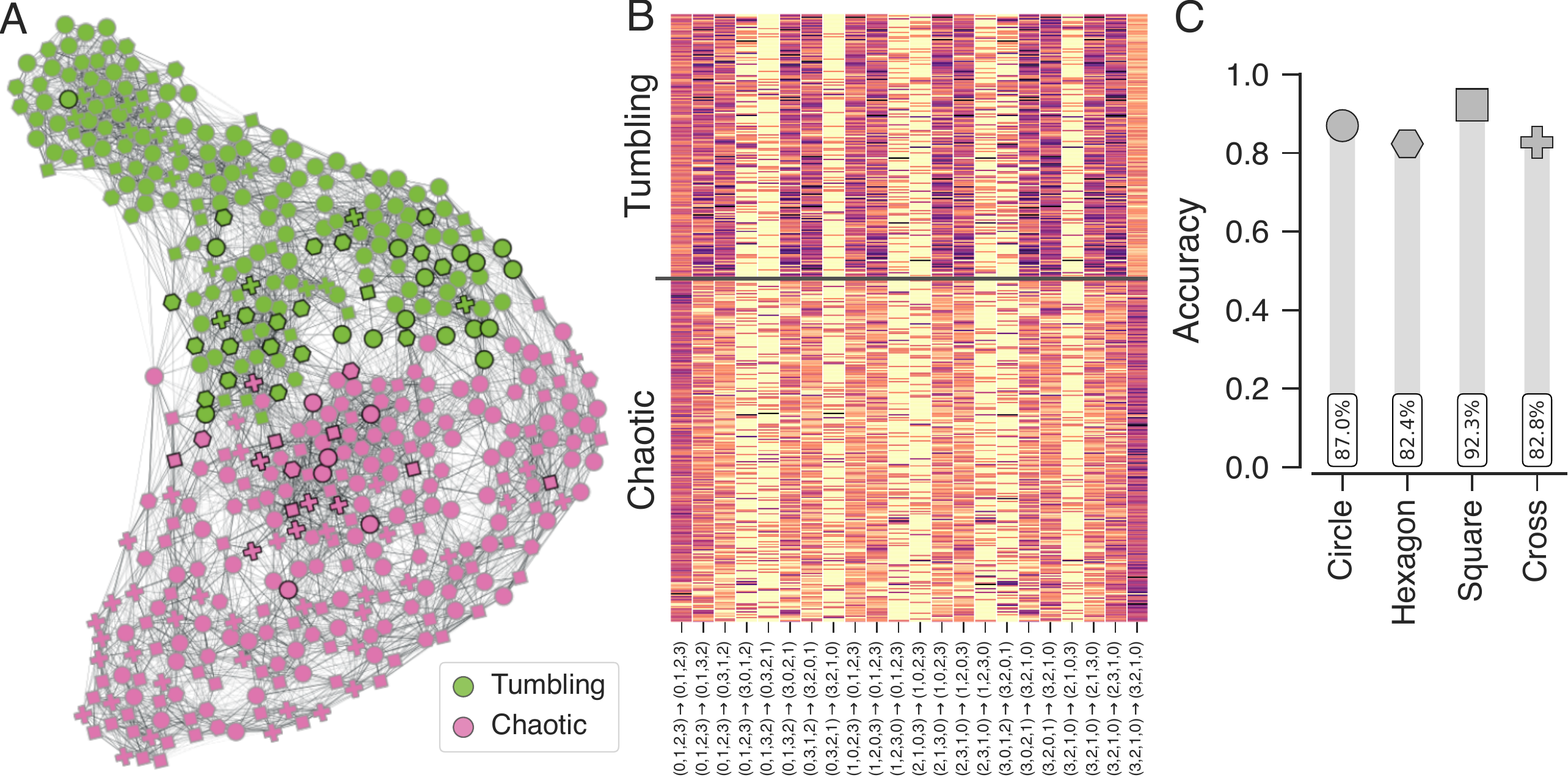}
    \caption{Modular structure of the similarity matrix obtained from ordinal-pattern transitions aligns with expert-labeled motion classes. (A) Similarity network of falling-paper trajectories with the two-module structure emerging from the infomap community detection algorithm highlighted in green and pink. Green nodes predominantly correspond to trajectories labeled as tumbling motion by experts, while pink nodes mostly represent trajectories classified as chaotic motion. Marker shapes denote paper geometry, and thicker black borders indicate trajectories (60 out of 441; 86.4\% accuracy) that disagree with expert classification. (B) Ordinal representation of falling-paper trajectories arranged according to the modular structure of the similarity network. Rows in the upper part of the matrix correspond to the module primarily comprising tumbling motion trajectories, while rows in the lower part correspond to chaotic motion. Darker shades represent higher probabilities for a given high-frequency transition between permutation patterns (informed below the columns) for each free fall in our dataset. (C) Classification accuracy by paper geometry. Square-shaped papers exhibit the highest accuracy (92\%), followed by circular (87\%), cross-shaped (83\%), and hexagonal (82\%) papers.}
\label{fig:2}
\end{figure*}

Figure~\ref{fig:2}A displays the optimal two-module partition of the similarity network, with nodes from each group colored in green and pink. The green module contains 193 nodes, most of which correspond to time series labeled as tumbling motion by experts, while the remaining 248 pink nodes predominantly represent trajectories labeled as chaotic motion. The modular structure of the similarity network thus closely aligns with the experts' classification, with only 60 of 441 time series assigned to a module that differs from the experts' label (21 misclassified as chaotic and 39 as tumbling). This yields an overall binary classification accuracy of $86.4\%$, approximately 7\% higher than that achieved using the dimensionless moment of inertia and the Reynolds number~\cite{howison2020large-scale}, and 5\% higher than results based on the complexity-entropy plane~\cite{pessa2022clustering}. Moreover, unlike Ref.~\cite{howison2020large-scale}, which relies on three-dimensional reconstructions of falling paper motion, our method operates solely on one-dimensional time series of the observed paper area. Notably, in contrast to both prior approaches, our technique allows motion types to emerge organically from the data, without requiring predefined feature space boundaries or a pre-fixed number of clusters. 

This improved performance reflects the quality of the ordinal representation obtained from the ordinal-pattern transitions. Indeed, by focusing only on a subset of transitions that consistently exhibit high frequencies across time series, Figure~\ref{fig:2}B shows that tumbling and chaotic motions display some distinct frequencies of ordinal-pattern transitions. Notably, self-transitions of the ascending and descending ordinal patterns [$(0,1,2,3)\to(0,1,2,3)$ and $(3,2,1,0)\to(3,2,1,0)$] are typically underrepresented in tumbling trajectories, most likely reflecting the greater number of local ups and downs (peaks and troughs) in observable area series of tumbling falls, leading to transitions between distinct permutation patterns that break the monotonic trends between these local extrema (see Figure~\ref{fig:1}{A}). In contrast, some transitions that are not connected to the persistent ups and downs of tumbling series [for instance, $(0,1,2,3)\to(0,1,3,2)$, $(0,1,2,3)\to(3,0,2,1)$, $(3,2,1,0)\to(2,1,3,0)$, and $(3,2,1,0)\to(2,3,1,0)$] tend to be overrepresented in chaotic trajectories, reflecting their smaller number of local extrema and even smoother observable area series associate with their swoops and aperiodic oscillations (as shown in Figure~\ref{fig:1}{A}). Nevertheless, it is evident that no individual transition suffices to clearly differentiate between chaotic and tumbling motions; rather, it is the collective structure embedded in the high-dimensional set of transitions that underlies the observed classification accuracy.

When disaggregating the results by paper geometry, Figure~\ref{fig:2}C shows that the performance of our approach varies across the paper formats. Square-shaped papers exhibit the highest accuracy (92\%), followed by circular (87\%), cross-shaped (83\%), and hexagonal (82\%) papers. Except for hexagonal papers --  for which Ref~\cite{pessa2022clustering} achieves a marginally higher accuracy -- our method consistently outperforms previous approaches. In particular, for cross-shaped papers, our approach attains an accuracy of 83\%, markedly surpassing the 64\% and 69\% reported in Refs.~\cite{howison2020large-scale} and \cite{pessa2022clustering}, respectively. It is worth noting that baseline classification strategies that assign the most common behavior to all falls (60\% of motions as chaotic) or use class frequencies to guide predictions produce substantially lower accuracies than our method. To further validate our findings, we apply our clustering approach to randomly shuffled time series, which results in a substantially larger number of clusters -- in stark contrast to the experts' binary labels -- and produces significantly higher minimum description lengths than those obtained from the original data. These results confirm that the observed two-module partition cannot be attributed to chance, but instead reflects the intrinsic geometry of the high-dimensional space defined by ordinal-pattern transitions.

\begin{figure*}[!t]
\centering
\includegraphics[width=0.5\linewidth]{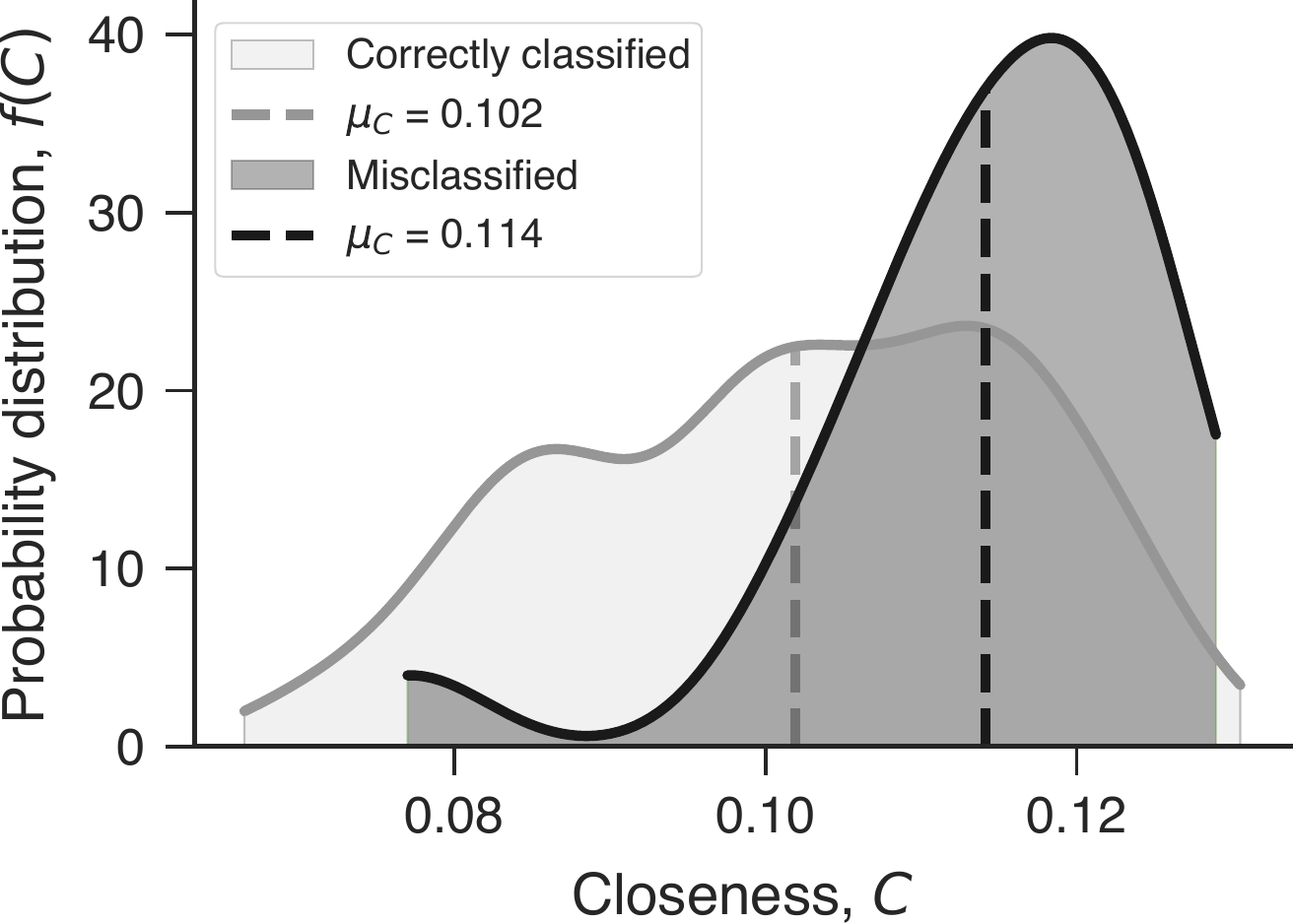}
   \caption{Misclassified trajectories tend to occupy more central positions in the similarity network. Probability distribution of closeness centrality ($C$) for trajectories correctly classified (light grey) and misclassified (dark grey) by our method. The distribution for misclassified trajectories is shifted toward higher values of $C$, indicating greater centrality. Vertical dashed lines denote the mean closeness centrality ($\mu_C$) for each group, with misclassified trajectories exhibiting significantly higher average centrality than those correctly classified.}
\label{fig:3}
\end{figure*}

We also investigate whether the 60 misclassified time series differ systematically from the 381 trajectories that our approach correctly classified as tumbling or chaotic, as visual inspection of Figure~\ref{fig:2}A suggests they tend to lie near the boundary between these two regimes. Specifically, we examine whether misclassified trajectories exhibit greater global similarity to other trajectories by comparing their weighted closeness centrality $C$ to that of the correctly classified ones. This centrality measure is defined as the reciprocal of the average shortest path distance from a node to all others, normalized by the number of reachable nodes, where the distance between nodes is given by the inverse of their similarity weight~\cite{newman2018networks}. As such, nodes strongly connected to many others have higher closeness values than those weakly connected. In our context, high closeness indicates trajectories that are broadly similar to many others in the network, potentially reflecting ambiguous or transitional dynamics that incorporate features of both chaotic and tumbling motions. Indeed, Figure~\ref{fig:3} shows that the distribution of closeness centrality for misclassified trajectories is significantly shifted toward higher values and is more narrowly concentrated compared to that of correctly classified trajectories, with average $C$ markedly greater in the misclassified set. These results thus suggest that misclassified trajectories likely correspond to motion patterns that are not easily separable by sharp regime boundaries -- such as short-term alternating episodes of tumbling and chaotic motions or some form of dynamical blending of both regimes.

We further verify the robustness of our findings by varying the number of nearest neighbors (from 10 to 100) used to create the UMAP network, as well as considering -- in addition to $d=4$ -- embedding dimensions $d=3$ and $d=5$. The results for $d=4$ and $d=5$ yield best partitions of the similarity network with two clusters that display essentially the same agreement levels with the experts' classification $(\approx\!86\%)$. For $d=3$, the number of clusters is more unstable, suggesting that ordinal transitions among the 6 possible ordinal patterns do not fully capture the falling paper dynamics.

\section{Discussion and conclusions}\label{sec:conclusion}

We have presented a data-driven approach for classifying the complex free-fall dynamics of small paper fragments by combining ordinal time series analysis with graph clustering on a network in which nodes represent individual trajectories and weighted edges encode their dynamical similarities. By characterizing the trajectories through the relative frequencies of 96 ordinal-pattern transitions extracted from observed area time series -- directly extracted from video recordings -- our method captures nuanced features of these movements without requiring full knowledge of the underlying three-dimensional dynamics. Importantly, our framework enables the motion classes to emerge organically from the data without predefined feature space boundaries or a pre-fixed number of clusters. 

Our results show that this symbolic encoding, paired with community detection on the similarity network, produces motion clusters that align closely with experts' visual classifications,  consistently outperforming previous methods~\cite{howison2020large-scale, pessa2022clustering}. The gains are particularly notable for cross-shaped papers, where our approach achieves an accuracy of 83\%, substantially surpassing the 64\% and 69\% reported in earlier studies -- a geometry that proved to be especially challenging due to its distinct dynamics and the likely limitations of low-dimensional projections in capturing its behavior~\cite{howison2020large-scale, pessa2022clustering}.

We have also verified that misclassified trajectories tend to occupy central positions in the similarity network generated by our clustering approach, exhibiting significantly higher closeness centrality than correctly classified ones. This indicates that such trajectories are broadly similar to many others in the dataset and likely reflect transitional or blended behaviors, such as falls alternating between tumbling and chaotic dynamics, or residing near bifurcation boundaries separating motion regimes. Thus, the similarity network not only supports accurate classification but also enables the identification of transitional dynamics that future studies might explore. 

Our study thus demonstrates that ordinal-pattern transitions offer a scalable framework for representing time series from complex empirical systems, enabling the construction of a similarity space that faithfully captures the structure of free-fall dynamics without requiring full physical reconstruction or prior modeling assumptions. We also note that our approach is not computationally expensive. The symbolization and clustering (repeated 1,000 times) of free falls take a few minutes to run on a standard personal computer, and thus do not represent a bottleneck when compared to the time-consuming nature associated with the experimental task involved in collecting the free fall data.

Taken together, these findings contribute to the long-standing challenge of understanding falling paper dynamics, exemplifying how symbolic dynamics combined with network science can reveal meaningful structure in complex physical data.

\section*{CRediT authorship contribution statement}
\noindent {\bf Angelo A. Flores:} Conceptualization, Methodology, Software, Investigation, Data Curation, Visualization, Writing - Original Draft, Writing - Review \& Editing. {\bf Leonardo G. J. M. Voltarelli:} Conceptualization, Methodology, Software, Investigation, Data Curation, Visualization, Writing - Original Draft, Writing - Review \& Editing. {\bf Andre S. Sunahara:} Conceptualization, Methodology, Software, Investigation, Data Curation, Visualization, Writing - Original Draft, Writing - Review \& Editing. {\bf Haroldo V. Ribeiro:} Conceptualization, Methodology, Software, Investigation, Data Curation, Visualization, Writing - Original Draft, Writing - Review \& Editing. {\bf Arthur A. B. Pessa:} Conceptualization, Methodology, Software, Investigation, Data Curation, Visualization, Writing - Original Draft, Writing - Review \& Editing.

\section*{Declaration of competing interest}
\noindent The authors declare that they have no known competing financial interests or personal relationships that could have appeared to influence the work reported in this paper.

\section*{Acknowledgments}
\noindent The authors acknowledge the support of Coordena\c{c}\~ao de Aperfei\c{c}oamento de Pessoal de N\'ivel Superior (CAPES) and the Conselho Nacional de Desenvolvimento Cient\'ifico e Tecnol\'ogico (CNPq -- Grant 303533/2021-8).

\section*{Data availability}
\noindent All code and data necessary to reproduce the results and figures in this work are available at the GitLab repository \url{https://gitlab.com/arthurpessa/ordinal-nets-free-fall}.

\bibliographystyle{elsarticle-num} 

\bibliography{refs.bib}

\begin{thebibliography}{10}
\expandafter\ifx\csname url\endcsname\relax
  \def\url#1{\texttt{#1}}\fi
\expandafter\ifx\csname urlprefix\endcsname\relax\def\urlprefix{URL }\fi
\expandafter\ifx\csname href\endcsname\relax
  \def\href#1#2{#2} \def\path#1{#1}\fi

\bibitem{blitzstein2019introduction}
J.~K. Blitzstein, J.~Hwang, Introduction to Probability, Second Edition,
  Chapman \& Hall/CRC Texts in Statistical Science, CRC Press, 2019.

\bibitem{bartos2024fair}
F.~Bartoš, A.~Sarafoglou, H.~R. Godmann, A.~Sahrani, D.~K. Leunk, P.~Y. Gui,
  D.~Voss, K.~Ullah, M.~J. Zoubek, F.~Nippold, F.~Aust, F.~F. Vieira, C.-G.
  Islam, A.~J. Zoubek, S.~Shabani, J.~Petter, I.~B. Roos, A.~Finnemann, A.~B.
  Lob, M.~F. Hoffstadt, J.~Nak, J.~de~Ron, K.~Derks, K.~Huth, S.~Terpstra,
  T.~Bastelica, M.~Matetovici, V.~L. Ott, A.~S. Zetea, K.~Karnbach, M.~C.
  Donzallaz, A.~John, R.~M. Moore, F.~Assion, R.~van Bork, T.~E. Leidinger,
  X.~Zhao, A.~K. Motaghi, T.~Pan, H.~Armstrong, T.~Peng, M.~Bialas, J.~Y.~C.
  Pang, B.~Fu, S.~Yang, X.~Lin, D.~Sleiffer, M.~Bognar, B.~Aczel, E.-J.
  Wagenmakers, Fair coins tend to land on the same side they started: Evidence
  from 350,757 flips, arXiv (2024).
\newblock \href {https://doi.org/10.48550/arXiv.2310.04153}
  {\path{doi:10.48550/arXiv.2310.04153}}.

\bibitem{zeng-yuan1985onthesensitive}
Y.~Zeng-yuan, Z.~Bin, On the sensitive dynamical system and the transition from
  the apparently deterministic process to the completely random process,
  Applied Mathematics and Mechanics 6~(3) (1985) 193--211.
\newblock \href {https://doi.org/10.1007/BF01895515}
  {\path{doi:10.1007/BF01895515}}.

\bibitem{diaconis2007dynamical}
P.~Diaconis, S.~Holmes, R.~Montgomery, Dynamical bias in the coin toss, SIAM
  Review 49~(2) (2007) 211--235.
\newblock \href {https://doi.org/10.1137/S0036144504446436}
  {\path{doi:10.1137/S0036144504446436}}.

\bibitem{maxwell1853particular}
J.~C. Maxwell, On a particular case of the descent of a heavy body in a
  resisting medium, Vol.~1 of Cambridge Library Collection - Physical Sciences,
  Cambridge University Press, 2011, p. 115–118.
\newblock \href {https://doi.org/10.1017/CBO9780511698095.008}
  {\path{doi:10.1017/CBO9780511698095.008}}.

\bibitem{tanabe1994behavior}
Y.~Tanabe, K.~Kaneko, Behavior of a falling paper, Physical Review Letters 73
  (1994) 1372--1375.
\newblock \href {https://doi.org/10.1103/PhysRevLett.73.1372}
  {\path{doi:10.1103/PhysRevLett.73.1372}}.

\bibitem{mahadevan1995comment}
L.~Mahadevan, H.~Aref, S.~W. Jones, Comment on ``{B}ehavior of a falling
  paper'', Physical Review Letters 75 (1995) 1420--1420.
\newblock \href {https://doi.org/10.1103/PhysRevLett.75.1420}
  {\path{doi:10.1103/PhysRevLett.75.1420}}.

\bibitem{field1997chaotic}
S.~B. Field, M.~Klaus, M.~G. Moore, F.~Nori, Chaotic dynamics of falling disks,
  Nature 388 (1997) 252--254.
\newblock \href {https://doi.org/10.1038/40817} {\path{doi:10.1038/40817}}.

\bibitem{belmonte1998fromflutter}
A.~Belmonte, H.~Eisenberg, E.~Moses, From flutter to tumble: {Inertial} drag
  and froude similarity in falling paper, Physical Review Letters 81 (1998)
  345--348.
\newblock \href {https://doi.org/10.1103/PhysRevLett.81.345}
  {\path{doi:10.1103/PhysRevLett.81.345}}.

\bibitem{mahadevan1999tumbling}
L.~Mahadevan, W.~S. Ryu, A.~D.~T. Samuel, Tumbling cards, Physics of Fluids
  11~(1) (1999) 1--3.
\newblock \href {https://doi.org/10.1063/1.869919}
  {\path{doi:10.1063/1.869919}}.

\bibitem{pesavento2004falling}
U.~Pesavento, Z.~J. Wang, Falling paper: {Navier-Stokes} solutions, model of
  fluid forces, and center of mass elevation, Physical Review Letters 93 (2004)
  144501.
\newblock \href {https://doi.org/10.1103/PhysRevLett.93.144501}
  {\path{doi:10.1103/PhysRevLett.93.144501}}.

\bibitem{andersen2005analysis}
A.~Andersen, U.~Pesavento, Z.~J. Wang, Analysis of transitions between
  fluttering, tumbling and steady descent of falling cards, Journal of Fluid
  Mechanics 541 (2005) 91–104.
\newblock \href {https://doi.org/10.1017/S0022112005005847}
  {\path{doi:10.1017/S0022112005005847}}.

\bibitem{howison2019physics}
T.~Howison, J.~Hughes, F.~Giardina, F.~Iida, Physics driven behavioural
  clustering of free-falling paper shapes, PLOS ONE 14~(6) (2019) e0217997.
\newblock \href {https://doi.org/10.1371/journal.pone.0217997}
  {\path{doi:10.1371/journal.pone.0217997}}.

\bibitem{howison2020large-scale}
T.~Howison, J.~Hughes, F.~Iida, Large-scale automated investigation of
  free-falling paper shapes via iterative physical experimentation, Nature
  Machine Intelligence 2 (2020) 68--75.
\newblock \href {https://doi.org/10.1038/s42256-019-0135-z}
  {\path{doi:10.1038/s42256-019-0135-z}}.

\bibitem{pessa2022clustering}
A.~A.~B. Pessa, M.~Perc, H.~V. Ribeiro, Clustering free-falling paper motion
  with complexity and entropy, EPL 138~(3) (2022) 30003.
\newblock \href {https://doi.org/10.1209/0295-5075/ac6bbb}
  {\path{doi:10.1209/0295-5075/ac6bbb}}.

\bibitem{willmarth1964steady}
W.~W. Willmarth, N.~E. Hawk, R.~L. Harvey, Steady and unsteady motions and
  wakes of freely falling disks, The Physics of Fluids 7~(2) (1964) 197--208.
\newblock \href {https://doi.org/10.1063/1.1711133}
  {\path{doi:10.1063/1.1711133}}.

\bibitem{rosso2007distinguishing}
O.~A. Rosso, H.~A. Larrondo, M.~T. Martin, A.~Plastino, M.~A. Fuentes,
  Distinguishing noise from chaos, Physical Review Letters 99 (2007) 154102.
\newblock \href {https://doi.org/10.1103/PhysRevLett.99.154102}
  {\path{doi:10.1103/PhysRevLett.99.154102}}.

\bibitem{bandt2002permutation}
C.~Bandt, B.~Pompe, Permutation entropy: {A} natural complexity measure for
  time series, {Physical Review Letters} 88 (2002) 174102.
\newblock \href {https://doi.org/10.1103/PhysRevLett.88.174102}
  {\path{doi:10.1103/PhysRevLett.88.174102}}.

\bibitem{small2013complex}
M.~Small, Complex networks from time series: {Capturing} dynamics, in: 2013
  {IEEE} International Symposium on Circuits and Systems ({ISCAS}2013), 2013,
  pp. 2509--2512.
\newblock \href {https://doi.org/10.1109/ISCAS.2013.6572389}
  {\path{doi:10.1109/ISCAS.2013.6572389}}.

\bibitem{mccullough2015time}
M.~McCullough, M.~Small, T.~Stemler, H.~H.-C. Iu, Time lagged ordinal partition
  networks for capturing dynamics of continuous dynamical systems, Chaos 25~(5)
  (2015) 053101.
\newblock \href {https://doi.org/10.1063/1.4919075}
  {\path{doi:10.1063/1.4919075}}.

\bibitem{pessa2019characterizing}
A.~A.~B. Pessa, H.~V. Ribeiro, Characterizing stochastic time series with
  ordinal networks, Physical Review E 100~(4) (2019) 042304.
\newblock \href {https://doi.org/10.1103/PhysRevE.100.042304}
  {\path{doi:10.1103/PhysRevE.100.042304}}.

\bibitem{olivares2020contrasting}
F.~Olivares, M.~Zanin, L.~Zunino, D.~P{\'e}rez, Contrasting chaotic with
  stochastic dynamics via ordinal transition networks, Chaos 30~(6) (2020)
  063101.
\newblock \href {https://doi.org/10.1063/1.5142500}
  {\path{doi:10.1063/1.5142500}}.

\bibitem{pessa2020mapping}
A.~A.~B. Pessa, H.~V. Ribeiro, Mapping images into ordinal networks, Physical
  Review E 102 (2020) 052312.
\newblock \href {https://doi.org/10.1103/PhysRevE.102.052312}
  {\path{doi:10.1103/PhysRevE.102.052312}}.

\bibitem{lee2021non}
E.~K. Lee, H.~Balasubramanian, A.~Tsolias, S.~U. Anakwe, M.~Medalla, K.~V.
  Shenoy, C.~Chandrasekaran, Non-linear dimensionality reduction on
  extracellular waveforms reveals cell type diversity in premotor cortex, Elife
  10 (2021) e67490.
\newblock \href {https://doi.org/10.7554/eLife.67490}
  {\path{doi:10.7554/eLife.67490}}.

\bibitem{sunahara2023universal}
A.~S. Sunahara, M.~Perc, H.~V. Ribeiro, Universal productivity patterns in
  research careers, Physical Review Research 5 (2023) 043203.
\newblock \href {https://doi.org/10.1103/PhysRevResearch.5.043203}
  {\path{doi:10.1103/PhysRevResearch.5.043203}}.

\bibitem{rosvall2008maps}
M.~Rosvall, C.~T. Bergstrom, Maps of random walks on complex networks reveal
  community structure, Proceedings of the National Academy of Sciences 105~(4)
  (2008) 1118--1123.
\newblock \href {https://doi.org/10.1073/pnas.0706851105}
  {\path{doi:10.1073/pnas.0706851105}}.

\bibitem{rosvall2009map}
M.~Rosvall, D.~Axelsson, C.~T. Bergstrom, The map equation, The European
  Physical Journal Special Topics 178~(1) (2009) 13--23.
\newblock \href {https://doi.org/10.1140/epjst/e2010-01179-1}
  {\path{doi:10.1140/epjst/e2010-01179-1}}.

\bibitem{howison2019github}
T.~Howison, {Falling-Paper}, {Available:}
  \url{https://github.com/th533/Falling-Paper}, {Accessed:} 03 Dec 2024 (2019).

\bibitem{cao2004detecting}
Y.~Cao, W.~wen Tung, J.~B. Gao, V.~A. Protopopescu, L.~M. Hively, Detecting
  dynamical changes in time series using the permutation entropy, Physical
  Review E 70 (2004) 046217.
\newblock \href {https://doi.org/10.1103/PhysRevE.70.046217}
  {\path{doi:10.1103/PhysRevE.70.046217}}.

\bibitem{amigo2008combinatorial}
J.~Amig{\'o}, S.~Zambrano, M.~A. Sanju{\'a}n, Combinatorial detection of
  determinism in noisy time series, EPL (Europhysics Letters) 83~(6) (2008)
  60005.
\newblock \href {https://doi.org/10.1209/0295-5075/83/60005}
  {\path{doi:10.1209/0295-5075/83/60005}}.

\bibitem{pessa2021ordpy}
A.~A.~B. Pessa, H.~V. Ribeiro, {ordpy}: {A} {Python} package for data analysis
  with permutation entropy and ordinal network methods, Chaos: An
  Interdisciplinary Journal of Nonlinear Science 31~(6) (2021) 063110.
\newblock \href {https://doi.org/10.1063/5.0049901}
  {\path{doi:10.1063/5.0049901}}.

\bibitem{mcinnes2018umap}
L.~McInnes, J.~Healy, J.~Melville, Umap: {Uniform} manifold approximation and
  projection for dimension reduction, arXiv (2018).
\newblock \href {https://doi.org/10.48550/arXiv.1802.03426}
  {\path{doi:10.48550/arXiv.1802.03426}}.

\bibitem{cunha2024shape}
L.~R. Cunha, A.~A. Pessa, R.~S. Mendes, Shape patterns in popularity series of
  video games, Chaos, Solitons \& Fractals 185 (2024) 115081.
\newblock \href {https://doi.org/10.1016/j.chaos.2024.115081}
  {\path{doi:10.1016/j.chaos.2024.115081}}.

\bibitem{lause2024theart}
J.~Lause, P.~Berens, D.~Kobak, The art of seeing the elephant in the room: {2D}
  embeddings of single-cell data do make sense, PLOS Computational Biology
  20~(10) (2024) 1--5.
\newblock \href {https://doi.org/10.1371/journal.pcbi.1012403}
  {\path{doi:10.1371/journal.pcbi.1012403}}.

\bibitem{mcinnes2018umap-software}
L.~McInnes, J.~Healy, N.~Saul, L.~Grossberger, Umap: {Uniform} manifold
  approximation and projection, The Journal of Open Source Software 3~(29)
  (2018) 861.
\newblock \href {https://doi.org/10.21105/joss.00861}
  {\path{doi:10.21105/joss.00861}}.

\bibitem{edler2022themap}
D.~Edler, A.~Holmgren, M.~Rosvall, \href{https://mapequation.org}{{The
  MapEquation software package}} (2022).
\newline\urlprefix\url{https://mapequation.org}

\bibitem{lancichinetti2009community}
A.~Lancichinetti, S.~Fortunato, Community detection algorithms: {A} comparative
  analysis, Physical Review E 80~(5) (2009) 056117.
\newblock \href {https://doi.org/10.1103/PhysRevE.80.056117}
  {\path{doi:10.1103/PhysRevE.80.056117}}.

\bibitem{fortunato2010community}
S.~Fortunato, Community detection in graphs, Physics Reports 486~(3-5) (2010)
  75--174.
\newblock \href {https://doi.org/10.1016/j.physrep.2009.11.002}
  {\path{doi:10.1016/j.physrep.2009.11.002}}.

\bibitem{fortunato2016community}
S.~Fortunato, D.~Hric, Community detection in networks: {A} user guide, Physics
  Reports 659 (2016) 1--44.
\newblock \href {https://doi.org/10.1016/j.physrep.2016.09.002}
  {\path{doi:10.1016/j.physrep.2016.09.002}}.

\bibitem{newman2018networks}
M.~Newman, Networks, 2nd Edition, Orford University Press, 2018.

\end{thebibliography}

\end{document}